# Speech Enhancement with Zero-Shot Model Selection

Ryandhimas E. Zezario*
Department of Computer Science
and Information Engineering
National Taiwan University
Taipei, Taiwan
ryandhimas@citi.sinica.edu.tw

Chiou-Shann Fuh
Department of Computer Science
and Information Engineering
National Taiwan University
Taipei, Taiwan
fuh@csie.ntu.edu.tw

Hsin-Min Wang
Institute of Information Science
Academia Sinica
Taipei, Taiwan
whm@iis.sinica.edu.tw

Yu Tsao
*Research Center for Information
Technology Innovation
Academia Sinica
Taipei, Taiwan
yu.tsao@citi.sinica.edu.tw

*Abstract*—Recent research on speech enhancement (SE) has seen the emergence of deep-learning-based methods. It is still a challenging task to determine the effective ways to increase the generalizability of SE under diverse test conditions. In this study, we combine zero-shot learning and ensemble learning to propose a zero-shot model selection (ZMOS) approach to increase the generalization of SE performance. The proposed approach is realized in the offline and online phases. The offline phase clusters the entire set of training data into multiple subsets and trains a specialized SE model (termed component SE model) with each subset. The online phase selects the most suitable component SE model to perform the enhancement. Furthermore, two selection strategies were developed: selection based on the quality score (QS) and selection based on the quality embedding (QE). Both QS and QE were obtained using a Quality-Net, a non-intrusive quality assessment network. Experimental results confirmed that the proposed ZMOS approach can achieve better performance in both seen and unseen noise types compared to the baseline systems and other model selection systems, which indicates the effectiveness of the proposed approach in providing robust SE performance.

## I. Introduction

Speech enhancement (SE) is an important front-end module for various speech-related applications, such as automatic speech recognition (ASR) [1–3], assistive listening [4–8], speech coding [9–10], and speaker recognition [11–12] systems. The primary aim of SE is to retrieve clean speech signals from noisy signals. With the emergence of deep learning algorithms, notable improvements in SE have been made over the traditional SE methods. Well-known examples include the fully connected neural network [13–14], deep denoising auto-encoder (DDAE) [15–17], convolutional neural network (CNN) [18–19], long short-term memory (LSTM) [20–21] and their combinations [22–24]. Despite past promising improvements, increasing the generalizability of deep-learning-based SE methods to unseen environments remains a critical research topic.

Zero-shot learning is a machine learning algorithm that has been proven to be capable of improving generalizability to unseen environments. This learning criterion has been successfully implemented in the field of image processing to recognize unseen objects with satisfactory performance [22–26]. In the field of speech processing, several attempts have been made to incorporate a zero-shot learning algorithm for robust performance [27–30]. For instance, in [29], speaker embedding was extracted and used as additional guidance to conduct noise reduction. In [30], noise embedding, namely the dynamic noise embedding, was extracted and used to characterize background noise information to develop more optimal noise reduction performance. However, most of the current zero-shot learning strategies rely on a similar fashion, where the generated latent representation is incorporated as an additional feature into the main task. In addition, due to the notable success of model selection approaches [31–32], we aim to use zero-shot learning as a model selection approach. To the best of our knowledge, no prior work has proposed the use of latent representations for model selection in speech enhancement tasks.

In this study, we propose a novel zero-shot model selection (ZMOS) approach for SE. The proposed approach combines zero-shot learning and ensemble learning to improve SE performance under any specific test condition and is implemented in two phases: offline and online. In the offline phase, we prepared multiple specialized SE models (termed component SE models). Each component SE model was trained to match the specific noisy condition. In the online phase, we selected the most suitable component SE model to enhance the test utterance. For the proposed approach, the effective clustering of the training data to train the multiple-component SE models in the offline phase and selecting the most suitable component SE model for a test utterance in the online phase are critical points. We propose to perform data clustering and model selection using a pre-trained Quality-Net [33]. A Quality-Net is a deep-learning-based non-intrusive quality assessment model. Given an utterance, the Quality-Net outputs a quality assessment score. Previous studies have shown that the Quality-Net can accurately predict the quality assessment score of an utterance.

Two types of data clustering and model selection strategies are developed: one is based on the quality score (QS), and the other is based on the quality embedding (QE); the corresponding approaches are termed as ZMOS-QS and ZMOS-QE, respectively. Both the QS and QE were estimated using the Quality-Net. Given an utterance, the QS is based on the output score of the Quality-Net, and the QE is based on the embedding vector of Quality-Net. In the offline phase, QS or QE was used to group the training data into several clusters. Each cluster was used to train a specialized SE model. A centroid vector was computed to represent each specialized SE model. In the online phase, the QS or QE of the test utterance is used to identify one cluster of training data, i.e., the corresponding component SE model. Finally, the selected SE model was used to perform the enhancement. Notably, the other reference neural network models can be used to prepare features for data clustering and model selection. The Quality-Net was chosen because the model was trained to predict the quality score, so it should possess useful speech information.

To evaluate the proposed zero-shot model selection (ZMOS) approach, we adopted the perceptual evaluation of speech quality (PESQ) [34] and short-time objective intelligibility (STOI) [35] objective evaluation metrics. Experimental results under both seen and unseen noisy conditions show that the proposed approach can achieve notable improvements compared with the baselines and other model selection approaches, thereby confirming the effectiveness of the proposed SE approach in providing robust enhancement performance.

## II. THE PROPOSED SYSTEMS

In this study, we propose two types of ZMOS strategies based on QS and QE. Both strategies share a similar concept by incorporating the Quality-Net as a reference model to extract quality features for performing the data clustering and model selection processes. In this section, we first review the Quality-Net model and introduce how to extract the QS and QE features with the Quality-Net. We then explain how to establish the ZMOS-QS and ZMOS-QE systems.

### A. Quality-Net

Quality-Net is a non-intrusive quality assessment neural network model trained with the aim of predicting utterance-level PESQ scores. As the length of the utterance varies, a bidirectional LSTM (BLSTM) is used to model the longer temporal information. In addition, to achieve a more accurate prediction score and mimic the human perceptive system, a conditional frame-wise constraint is introduced to train the model. Accordingly, the objective function of the Quality-Net is derived as follows

$$O = \frac{1}{N}\sum_{n=1}^{N}[(Q_n - \hat{Q}_n)^2 + \frac{\alpha(Q_n)}{L(U_n)}\sum_{l=1}^{L(U_n)}(Q_n - q_{n,l})^2] \quad (1)$$

where $N$ and $L(u_n)$ indicate the number of training utterances and the number of frames of the $n$-th utterance, respectively; $Q_n$ and $\hat{Q}_n$ indicate the true and predicted PESQ scores, respectively; and $q_{n,l}$ and $\alpha(Q_n)$ indicate the estimated frame-level quality of the $l$-th frame of utterance $n$ and weighting factor, respectively. Finally, given a noisy input $y_n$, the Quality-Net equation can be derived as follows:

$$\hat{Q}_n = QualityNet(y_n), \quad (2)$$

where $QualityNet(.)$ denotes the PESQ prediction function.

In our previous studies [32–33], we have confirmed the high prediction capability of the Quality-Net. We believe that both the output scores and latent representations of the Quality-Net provide useful information for determining the quality of given speech. This was the main motivation for this study.

### B. The Proposed System I: ZMOS-QS

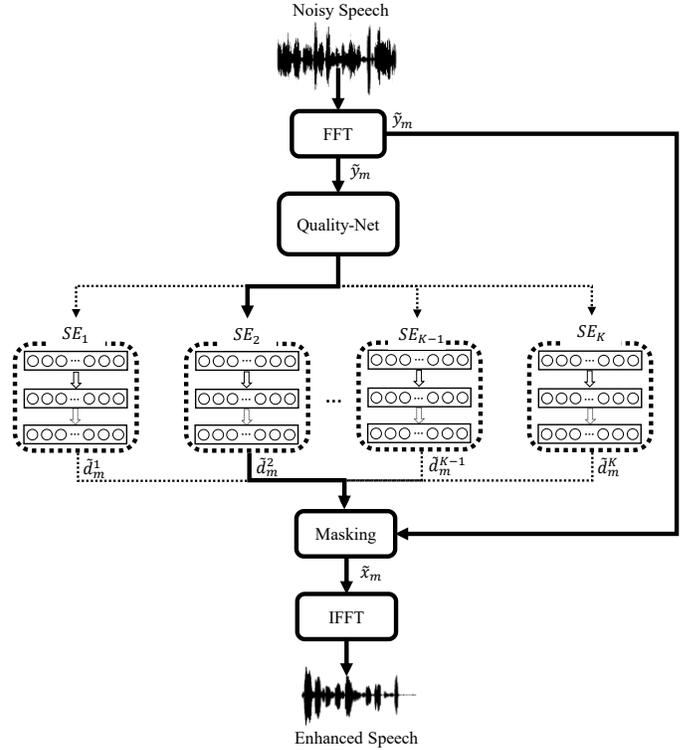

Fig. 1: The architecture of the ZMOS-QS approach.

The overall system architecture of the ZMOS-QS is shown in Fig. 1. In the training stage, we first apply the short-time Fourier transform (STFT) to convert speech waveforms into spectral features. With the paired spectral features, $\mathbf{Z}=[\mathbf{X}, \mathbf{Y}]$, which are formed by noisy spectral features $\mathbf{Y}$ and clean spectral features $\mathbf{X}$, PESQ scores are computed. They are used as a reference to cluster the entire set of training data into several subsets: $\{\mathbf{Z}_1, ..., \mathbf{Z}_t, ... \mathbf{Z}_T\}$, where $\mathbf{Z}_t$ is the $t$-th subset of paired training data, and $T$ is the total number of subsets. Based on the $T$ subsets of the training data, we then estimate the $T$-component SE models with an ideal ratio mask (IRM) [36] in the log domain as the training target criterion:

$$\begin{aligned}\mathbf{D}_1 &= F_1(\mathbf{Y}_1), \\ &\dots \\ \mathbf{D}_t &= F_t(\mathbf{Y}_t), \quad (3) \\ &\dots \\ \mathbf{D}_T &= F_T(\mathbf{Y}_T),\end{aligned}$$

where $\mathbf{Y}_t$, $\mathbf{D}_t$ and $F_t$ are the input, output, and transformation, respectively, of the $t$-th SE model.

In ZMOS-QS, the training data are clustered based on their PESQ scores predicted by the Quality-Net. Specifically, the PESQ scores were ranked. The training utterances with similar PESQ scores were grouped into a subset for training the corresponding component SE model. The average PESQ score for each subset was computed.

In the testing phase, given a noisy speech with the spectral feature $\tilde{\mathbf{y}}$, its PESQ is first estimated by the Quality-Net. Then, the enhancement is carried out by $\tilde{d} = F_t(\tilde{y})$, when $QualityNet(\tilde{y})$ is closest to the average PESQ score of the $t$-th component SE model and enhanced spectral feature $\tilde{x} = F_M(\tilde{d}, \tilde{y})$, where $F_M$ is the masking function. Finally, an inverse STFT is applied to reconstruct the enhanced speech waveform using enhanced spectral features, where the phase from the noisy speech is used.

*C. The Proposed System I: ZMOS-QE*

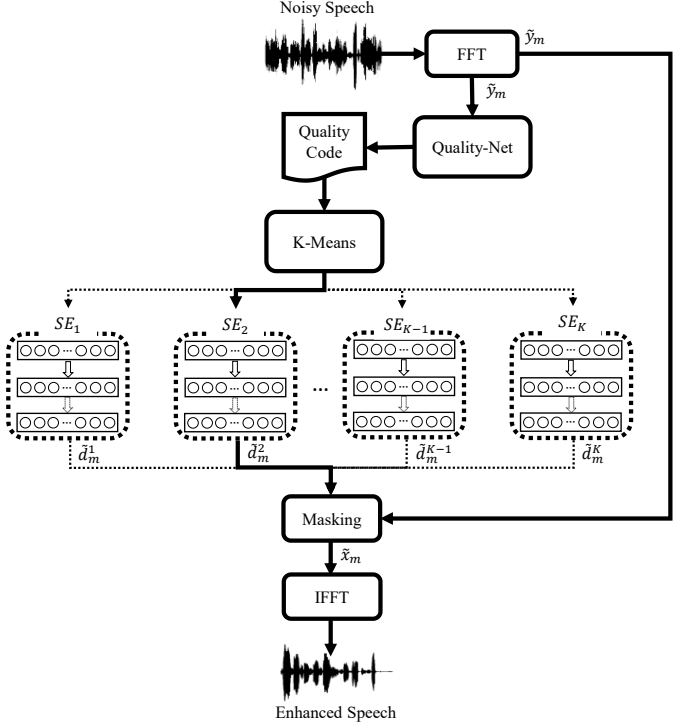

Fig. 2: The architecture of the ZMOS-QE approach.

ZMOS-QE adopts a similar idea to ZMOS-QS. Instead of QS, ZMOS-QE uses the latent representations of the Quality-Net to perform the data clustering and model selection, as shown in Fig. 2. In the training phase, given noisy spectral features, $Y = [y_1, ..., y_n ..., y_N]$, where $N$ is the total number of frames, a set of QE features, $Q = [q_1, ..., q_n ..., q_N]$, is extracted. Next, by applying the K-means algorithm to the entire set of QE features, we can cluster the QE features into $T$ clusters. Accordingly, the training data can be divided into $T$ subsets, $\{Z_1, ..., Z_t, ... Z_T\}$, represented by $T$ centroid QE vectors, $V = [v_1, ..., v_t, ... v_T]$, respectively. Then, we prepared $T$-component SE models, as shown in Eq. 3.

In the testing stage, given a noisy speech with spectral features, $\tilde{y}$, we first compute the QE feature, $\tilde{q}$, using the Quality-Net. Then, we calculate the distance between $\tilde{q}$ and each of the centroid QE features in $[v_1, ..., v_t, ... v_T]$. Next, we perform SE by $\tilde{d} = F_t(\tilde{y})$ if $v_t$ is closest to $\tilde{q}$ and obtain the enhanced spectral feature $F_M(\tilde{d}, \tilde{y})$. With the enhanced spectral feature, $\tilde{x}$, we can obtain the enhanced waveform by applying ISTFT along with the phase from the noisy speech.

## III. EXPERIMENTS

In this section, we first present the experimental setup, including the dataset preparation and the neural network model architectures. Next, we present the experimental results of ZMOS-QS and ZMOS-QE and discuss our findings.

*A. Experimental Setup*

We adopted the Wall Street Journal (WSJ) [37] dataset to evaluate the proposed ZMOS-QS and ZMOS-QE approaches. The WSJ dataset consists of 37,416 training and 330 test utterances recorded at a 16-kHz sampling rate. We prepared the noisy training utterances by injecting 100 types of stationary and non-stationary noises [38] into the WSJ training utterances at 31 signal-to-noise ratio (SNR) levels ranging from 20 to -10 dB with a step of 1 dB. For the test data, we prepared the noisy utterances by injecting two seen (white and engine noises) and two unseen (car and street noises) noise types at five SNR levels (-10, -5, 0, 5, and 10 dB). With a Hamming window of 32 ms and a hop size of 16 ms, a 512-point STFT was performed on the training and test utterances to extract 257-dimensional log-power spectra features.

We compared the proposed approaches with a CNN-based baseline system. The CNN model consisted of 12 convolutional layers, followed by a fully connected layer consisting of 128 neurons. Each convolutional layer contains four channels {16, 32, 64, and 128}. Each channel has three types of strides: {1, 1, 3}. The entire set of training utterances was used to train the CNN-based baseline. The component SE models in the ZMOS-QS and ZMOS-QE were implemented based on the same CNN architecture for a fair comparison. The training data were first divided into several subsets, with each subset used to train a component SE model. In this study, we divided the entire set of training data into four clusters. Therefore, there are four component SE models.

We used the standardized PESQ and STOI scores to evaluate the proposed ZMOS-QS and ZMOS-QE approaches. PESQ was used to evaluate the quality of speech, with a score ranging from -0.5 to 4.5. STOI was designed to evaluate the intelligibility of speech, with a score ranging from 0 to 1. Higher PESQ and STOI scores indicate that the enhanced speech has better speech quality and intelligibility, respectively.

*B. Objective Evaluation Results*

The average PESQ and STOI scores of unprocessed noisy speech, enhanced speech by CNN baseline, ZMOS-QS, and ZMOS-QE under white and engine noise types are shown in Tables 1. These two noise types were seen in the training. For comparison, we implemented and tested performance using another model selection method, specialized Speech Enhancement Model Selection (SSEMS) [32], in which the component models are trained to learn gender and signal-to-noise-ratio (SNR) information instead of the data-driven approach used in ZMOS. From Tables 1, we can note that both ZMOS-QS and ZMOS-QE achieve notably better PESQ and STOI scores than the unprocessed noisy speech, the baseline CNN system, and the SSEMS in both stationary and non-stationary noisy environments.

Table 1. PESQ and STOI comparison of Noisy, CNN, SSEMS, ZMOS-QS, ZMOS-QE systems under seen noise conditions (white and engine noises).

|        | PESQ | STOI |
|--------|------|------|
| Noisy  | 2.01 | 0.77 |
| CNN    | 2.42 | 0.78 |
| SSEMS [32] | 2.43 | 0.78 |
| ZMOS-QS | 2.46 | 0.79 |
| ZMOS-QE | 2.52 | 0.79 |

Table 2. PESQ and STOI comparison of Noisy, CNN, SSEMS, ZMOS-QS, ZMOS-QE systems under unseen noise conditions (car and street noises).

|        | PESQ | STOI |
|--------|------|------|
| Noisy  | 1.71 | 0.68 |
| CNN    | 2.49 | 0.80 |
| SSEMS [32] | 2.53 | 0.80 |
| ZMOS-QS | 2.51 | 0.81 |
| ZMOS-QE | 2.57 | 0.80 |

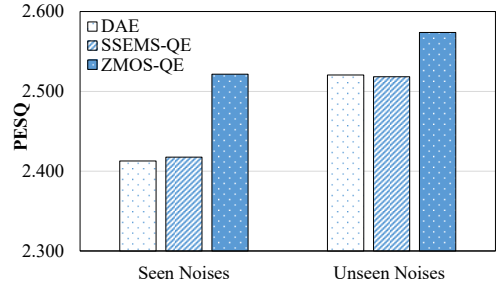

Fig.3: PESQ comparison of DAE, SSEMS-QE, and ZMOS-QE under seen and unseen conditions.

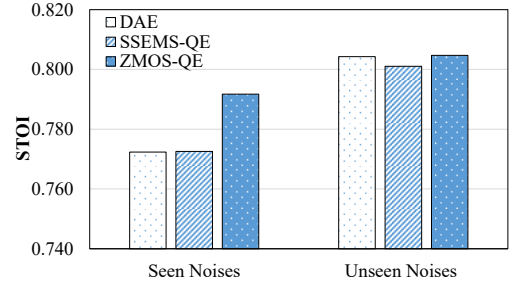

Fig.4: STOI comparison of DAE, SSEMS-QE, and ZMOS-QE under seen and unseen conditions.

Table 2 shows the average PESQ and STOI scores of the unprocessed noisy speech, the enhanced speech by CNN baseline, SSEMS, ZMOS-QS, and ZMOS-QE for car and street noise types. These two noise types were not observed during the training. From Table 2, we can again note that ZMOS-QE achieves considerably better performance compared to the other systems. The ZMOS-QS achieved better performance than the baseline systems and comparable performance with the SSEMS systems. Overall, the results confirm the effectiveness of the proposed approach for robust speech enhancement (SE) performance.

*C. Model Selection Analysis*

In the previous section, we demonstrated the effectiveness of the proposed method for noise reduction. In particular, we demonstrated the effectiveness of using latent representations to develop the component models and perform the model selection. Based on the notable performances achieved by ZMOS-QE, we conducted additional evaluations, where the comparative systems adopted the same component models as those used in ZMOS-QE but different model selection strategies.

Two other systems, namely the auto-encoder-based approach DAE [31] and SSEMS-QE [32], were developed. The DAE selects the best candidate based on the reconstruction error of the auto-encoder. Meanwhile, SSEMS-QE selects the best candidate based on the highest PESQ score given several component models. In contrast to the original SSEMS, SSEMS-QE adopted the quality embedding-based component models as those used in the ZMOS-QE approach. As shown in Figs. 3 and. 4, the proposed ZMOS-QE consistently overcomes the other selection methods in terms of PESQ and STOI scores under seen and unseen noises. Interestingly, unlike the other selection methods that require computing all possible component models to select the most suitable model, our proposed method can use only one utterance to identify the best fit model. Therefore, it can reduce the computational cost but yet still maintains better enhancement performances.

*D. Spectrogram Analysis*

In addition to the objective evaluations, we present the spectrograms to visualize the processed speech. Fig. 5 shows the spectrograms of a clean utterance (top left), corresponding noisy utterance at 0 dB SNR under car-noise (top right), enhanced speech by the CNN baseline (bottom left), and enhanced speech by ZMOS-QE (bottom right). We present the resulting spectrogram of ZMOS-QE only because ZMOS-QE has consistently achieved more effective reduction performance. From Fig. 5, we can confirm the effectiveness of the CNN baseline for SE. The proposed ZMOS-QE model can yield even better noise reduction results and recover the speech more accurately, compared with the CNN baseline, as seen in the red box; the speech processed by ZMOS-QE retains more detailed speech information than the CNN baseline.

## IV. CONCLUSIONS

In this study, we proposed two zero-shot model selection approaches for SE: ZMOS-QS and ZMOS-QE. The proposed approaches were derived based on zero-shot learning and ensemble learning. The quality score and embedding from the Quality-Net were used to perform data clustering and model selection. Experimental results confirmed that the proposed approaches effectively improve the SE performance of the baseline system, based on which the proposed approaches are built. To the best of our knowledge, this work is the first attempt to perform zero-shot learning as a model selection for SE and has improved the performance. In the future, we will explore the applicability of the proposed ZMOS approaches in other speech-processing tasks, such as dereverberation or cross-corpus SE tasks.

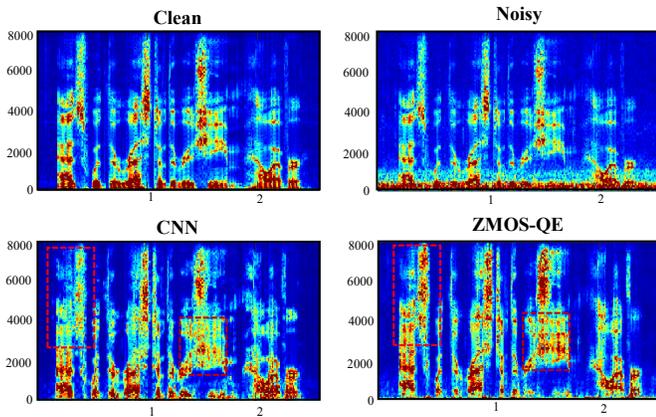

Fig.5: Spectrograms of a clean utterance (Clean), along with its noisy version (car noise at 0 dB SNR) (Noisy), and the CNN baseline and ZMOS-QE enhanced ones.